\documentstyle [11pt]{article}

\begin{document}

\title{{\bf Branching Space-time analysis of the GHZ theorem}\thanks{To 
be published in {\it Foundations of Physics} 1996.}} 
\author{{\em Nuel Belnap \thanks{E-mail: belnap+@pitt.edu}} 
\\ Department of Philosophy\\ 
University of Pittsburgh\\
$\ $
\\  {\em L\'aszl\'o E. Szab\'o} \thanks{Supported by OTKA 
Foundation, No: T -015606 and T-017176} \thanks{E-mail: 
leszabo@ludens.elte.hu}
\\Institute 
for Theoretical Physics\\ E\"otv\"os University, Budapest}
\maketitle
\newtheorem{D}{Definition}[section]
\newtheorem{T}{Theorem}[section]

\begin{abstract}

Greenberger, Horne, Shimony and Zeilinger gave a
new version of the Bell theorem without using inequalities
(probabilities). Mermin summarized it concisely; but Bohm and Hiley 
criticized Mermin's proof from contextualists' point of view.
 Using the Branching Space-time language, in this paper a proof will  
be given
that is free of  these difficulties. At the same time we will also
clarify the limits of the validity of the theorem when it is taken as
a proof that quantum mechanics is not compatible with a  
deterministic
world nor with a world that permits correlated space-related events
without a common cause. 
\end{abstract} 
\section{ Greenberger, Horne, Shimony, Zeilinger and Mermin }
\label{sec-GHSZ}  

Greenberger, Horne, Shimony and Zeilinger (1990) developed a proof 
of the Bell theorem without using inequalities. Following Mermin, one 
can extract from their ideas not only a simple Kochen-Specker-like 
illustration that there are  quantities that have no values  
independently 
of the measurements, but also some simple Bell-EPR like  results 
emphasizing locality (spatio-temporal) principles.

In the GHSZ example we consider three spin-half particles originated 
in a gedanken
spin-conserving decay flying apart along three different straight
lines in the horizontal plane. Let us denote the spin of particle $ i
$ along its direction of motion by $\frac{1}{2}\hbar \sigma ^{i}_{z}
$, the spin along the vertical direction by $\frac{1}{2}\hbar \sigma
^{i}_{x} $ and the spin along the horizontal direction orthogonal to
the trajectory by $\frac{1}{2}\hbar \sigma ^{i}_{y} $. Assume that
the quantum state of the three-particle system is
\begin{eqnarray}
\Psi =\frac{1}{\sqrt{2}}\left( \left| 1^{1}_{z} \right\rangle \otimes  
\left| 1^{2}_{z} \right\rangle \otimes \left| 1^{3}_{z} \right\rangle - 
\left| -1^{1}_{z} \right\rangle \otimes  \left|- 1^{2}_{z} \right\rangle 
\otimes \left| -1^{3}_{z} \right\rangle \right) 
\end{eqnarray}
where $\sigma ^{i}_{z}\left| 1^{i}_{z}\right\rangle =  \left| 
1^{i}_{z}\right\rangle $ and $\sigma ^{i}_{z}\left|- 
1^{i}_{z}\right\rangle =  \left|- 1^{i}_{z}\right\rangle $. In this state, 
the possible measurement results can be
$R(\sigma ^{a}_{x})=\pm 1 $  and $R(\sigma ^{a}_{y})=\pm 1 $  for 
$a=1,2,3$. 
Consider now the following operators:
\begin{eqnarray}
\Omega _{1}& = &\sigma ^{1}_{x}\sigma ^{2}_{y}\sigma ^{3}_{y}     
\nonumber  \\
\Omega _{2}& = &\sigma ^{1}_{y}\sigma ^{2}_{x}\sigma ^{3}_{y} 
\\
\Omega _{3}& = &\sigma ^{1}_{y}\sigma ^{2}_{y}\sigma ^{3}_{x}     
\nonumber \\
\Omega _{4}& = &\sigma ^{1}_{x}\sigma ^{2}_{x}\sigma 
^{3}_{x}\nonumber.
\end{eqnarray}
The only known way to measure these quantities is to perform the 
measurements $\sigma ^{1}_{x},\sigma ^{2}_{y},\sigma ^{3}_{y}$ 
and $\sigma ^{1}_{y},\sigma ^{2}_{x},\sigma ^{3}_{y},$ etc., and to 
take the product of the correponding results. Of course, the  
quantum 
state $\Psi $ does not determine the measurement results $R(\sigma 
^{1}_{x}), R(\sigma ^{2}_{y}), R(\sigma ^{3}_{y}), R(\sigma 
^{1}_{y}), R(\sigma ^{2}_{x}), R(\sigma ^{3}_{y}), \ldots\ .$ But we 
can know in advance each $R(\Omega_{i})$!   The state $\Psi $ is an 
eigenstate of $\Omega _{1},\Omega _{2}$ and $\Omega _{3}$ with 
eigenvalue 1. $\Psi $ is an eigenstate of $\Omega _{4}$, too, with 
eigenvalue $-1$. Consequently,  in any measurement of these  
quantities 
the results are 
\begin{eqnarray}
\label{constraint} 
R\left( \Omega _{1}\right)= R\left(
\Omega _{2}\right)= R\left( \Omega _{3} \right)=- R\left( \Omega 
_{4}\right)=1
\end{eqnarray}
which implies a constraint on the measurement results for the
spin-components of the three separated particles. This constraint
provides, what we shall call an {\em inconsistency-type correlation} 
among the outcomes at the three stations; that is, there are {\em 
combinations} of outcomes that are not possible, even though each 
{\em individual} outcome is possible.

The measurement results $R(\Omega_{i})$ are fixed in advance.    
That 
is why, according to the minimal understanding of  a ``value"  
assigned 
to a quantum observable, one may say that the values of 
$\Omega_{1},\ \ldots \Omega_{4}$  are
\begin{eqnarray}
\label{value-constraint} 
V\left( \Omega _{1}\right)= V\left(
\Omega _{2}\right)= V\left( \Omega _{3} \right)=- V\left( \Omega 
_{4}\right)=1
\end{eqnarray}
Consider now the product of these operators
\begin{eqnarray}
\Omega= \Omega_{1}\Omega_{2}\Omega_{3}\Omega_{4}.
\end{eqnarray}
 $\Psi$ is an eigenstate of $\Omega $ too with eigenvalue $-1$. This  
fact 
is consistent with
\begin{eqnarray}
\label{vvan}
V(\Omega_{1}) V(\Omega_{2}) V(\Omega_{}) V(\Omega_{4})=-1,
\end{eqnarray}
that is,
\begin{eqnarray}
\label{vvan9}
V(\sigma^{1}_{x}\sigma^{1}_{y}\sigma^{1}_{y}) 
V(\sigma^{1}_{y}\sigma^{1}_{x}\sigma^{1}_{y}) 
V(\sigma^{1}_{y}\sigma^{1}_{y}\sigma^{1}_{x}) 
V(\sigma^{1}_{x}\sigma^{1}_{x}\sigma^{1}_{x})=-1.
\end{eqnarray}
And the same holds for the measurement results
\begin{eqnarray}
\label{rvan9}
R(\sigma^{1}_{x}\sigma^{1}_{y}\sigma^{1}_{y}) 
R(\sigma^{1}_{y}\sigma^{1}_{x}\sigma^{1}_{y}) 
R(\sigma^{1}_{y}\sigma^{1}_{y}\sigma^{1}_{x}) 
R(\sigma^{1}_{x}\sigma^{1}_{x}\sigma^{1}_{x})=-1.
\end{eqnarray}

Until now we did not  do anything beyond standard quantum 
mechanics. Here we come, however, to a Kochen-Specker-type 
argument. Let us make an additional assumption and see whether it 
goes through or not: Assume that we can assign values not only to  
the 
products $\sigma^{1}_{x}\sigma^{1}_{y}\sigma^{1}_{y},\ 
\sigma^{1}_{y}\sigma^{1}_{x}\sigma^{1}_{y},\ \ldots $, but also to 
the spin-operators $\sigma^{1}_{x},\ \sigma^{1}_{y},\ 
\sigma^{2}_{x},\ \sigma^{2}_{y},\ \sigma^{3}_{x},\ 
\sigma^{3}_{y}$  themselves, such that (\ref{vvan9}) can be written  
as
\begin{eqnarray}
\label{vimpossible}
V(\sigma^{1}_{x})V(\sigma^{2}_{y})V(\sigma^{3}_{y}) 
V(\sigma^{1}_{y})V(\sigma^{2}_{x})V(\sigma^{3}_{y})  \nonumber 
\\
V(\sigma^{1}_{y})V(\sigma^{2}_{y})V(\sigma^{3}_{x}) 
V(\sigma^{1}_{x})V(\sigma^{2}_{x})V(\sigma^{3}_{x})=-1.
\end{eqnarray}
This is, however, impossible, because each $V(\sigma^{a}_{j})$ 
appears twice, so whatever the values $V(\sigma^{a}_{j})$ are, the  
left hand 
side is a positive number, instead of $-1$. In this 
way, one finds evidence against  the idea of pre-existing values for  
the 
quantities $\sigma^{a}_{j}$.

From the Kochen-Specker point of view locality considerations  
are absent or not so important. From the EPR point of view, one  
takes  
into account that the various measurements are distant, and finds  
evidence that in spite of the correlations among them, there are no  
``elements of reality"  corresponding to the values $V\left( \sigma 
^{a}_{i} \right) $. One uses the term ``element of reality" partly 
because one thinks of a pre-settled value of a physical quantity 
as corresponding to an objective feature of our world, a feature  
that exists independently of the measurement that reveals it.  
Another 
reason is  that $V\left( \sigma ^{a}_{i} \right) $ seems to satisfy the 
often-quoted EPR reality criterion: 
{\em ``If, without in any way disturbing a system, we can predict  
with certainty the value of a physical quantity, then there exists an  
element of physical reality corresponding to this physical quantity."} 
By  virtue of the spatial separation of the three measurements, one  
can  
predict  any of $V\left( \sigma ^{a}_{i} \right), $ without apparent 
disturbance of  the system, by 
carrying out two suitable measurements on the other two particles.  
(See Mermin (1990a)). 

Since the values of the spin-components cannot be prearranged in 
advance before the measurements, the {\it individual } experiments  
cannot merely reveal values settled in advance. In this sense one  
can 
use  the GHSZ example in the Bell-EPR-common cause context, too:
The non-existence of prearranged values/measurement outcomes  
seems to  imply the non-existence of a {\em common cause} that  
can 
account for all the correlations provided by the constraint 
(\ref{constraint}).  
In the remainder of this paper we will be discussing the GHSZ  
example  
in this context.  In order to establish the correctness of the  
implication 
to no-common-cause, we use language that is more rigorous than  
customary.
In this way our discussion will not only provide a more solid  
nonlocality theorem, but will avoid the 
vulnerable points about which the contextualists' complaints have  
been made.

\section{ Contextualists' criticism} 

From the contextualists' point of view the above proof is not 
completely acceptable. Their critique is focused in the following 
remarks:
\begin{itemize}
\item From a contextualist point of view any Kochen-Specker-type 
argument is vacuous. The disagreement (\ref{vimpossible}), for 
example, proves the non-existence of values, the existence of  
which 
has never been assumed by a contextual theory, such as the Bohm 
mechanics. The "beables", if they exist, determine the result  of  
each 
individual measurement {\em operation}.  But these results are not 
present before the measurement operations have been completed.

\item  The Bell-EPR or  the Bell-EPR-common cause context 
concerns the question of what we can tell about the measurement  
results 
$$R(\sigma^{1}_{x}), \ R(\sigma^{1}_{y}), \ldots 
R(\sigma^{1}_{x}).$$ To avoid a contradiction analogous to 
(\ref{vimpossible}),
\begin{eqnarray}
\label{rimpossible}
\underbrace{ 
R(\sigma^{1}_{x})R(\sigma^{2}_{y})R(\sigma^{3}_{y})}_{1}
\underbrace{ 
R(\sigma^{1}_{y})R(\sigma^{2}_{x})R(\sigma^{3}_{y})}_{1}   
\nonumber \\
\underbrace{ 
R(\sigma^{1}_{y})R(\sigma^{2}_{y})R(\sigma^{3}_{x}) 
}_{1}\underbrace{ 
R(\sigma^{1}_{x})R(\sigma^{2}_{x})R(\sigma^{3}_{x})}_{-1} =-1,
\end{eqnarray}
the measurement results must be {\em context dependent} in the  
sense that, 
for example,  $ R(\sigma^{1}_{x})$ cannot be the same in the first 
place, when  $\sigma^{1}_{x}$  is assumed to be measured along  
with 
$\sigma^{2}_{y}$ and $\sigma^{3}_{y}$, as it is in the second place  
when 
it is  measured with $\sigma^{2}_{x}$  and $\sigma^{3}_{x}$ (see 
Bohm and Hiley 1993). So, it would be better to use a special index,  
to 
express the context in which the measurement is completed:
\begin{eqnarray}
\label{rpossible}
\underbrace{ 
R_{xyy}(\sigma^{1}_{x})R_{xyy}(\sigma^{2}_{y})R_{xyy}(\sigma^
{3}_{y})}_{1}\underbrace{ 
R_{yxy}(\sigma^{1}_{y})R_{yxy}(\sigma^{2}_{x})R_{yxy}(\sigma^
{3}_{y})}_{1}  \nonumber  \\
\underbrace{ 
R_{yyx}(\sigma^{1}_{y})R_{yyx}(\sigma^{2}_{y})R_{yyx}(\sigma^
{3}_{x}) }_{1}\underbrace{ 
R_{xxx}(\sigma^{1}_{x})R_{xxx}(\sigma^{2}_{x})R_{xxx}(\sigma^
{3}_{x})}_{-1} =-1,
\end{eqnarray}
Such an expression is not contradictory at all. 

One might argue that $R_{xyy}(\sigma^{1}_{x})$, for example, 
should  
be the same 
as $R_{xxx}(\sigma^{1}_{x})$, because of the spatial separation of 
the three stations. There is no influence on the measurement at 
station 1 of the choices among the possible measurements at the  
other two 
stations.  But this is irrelevant. It is a matter of fact that 
$R_{xyy}(\sigma^{1}_{x})$ differs from 
$R_{xxx}(\sigma^{1}_{x})$. These two measurements belong to 
different runs of the experiment.
\end{itemize}

We agree that both of the above remarks  are relevant, but we  
think that this is not the last word. In the following sections we will 
present a reformulation of Mermin version 
of the GHSZ thought-experiment. It will avoid vague notions such as 
``value of a quantity'', concentrating on the Bell-EPR-common cause 
context only.  
Instead,
\begin{itemize}
\begin{enumerate}
\item we will reformulate the problem using only some elementary
objects, such as events and
their causal relations, 

\item in this language we will describe exactly what
it could mean to say that there is no common-cause explanation of
inconsistency-type correlations among separated spin  
measurements,

\item and we will refer to the observed correlations among the 
separated spin measurements results only in such a 
way that everything will have a correct empirical meaning.
\end{enumerate}
\end{itemize}

On this purified basis, we will give an airtight proof that in
a certain limited sense, the GHSZ thought-experiment does in fact
imply a violation of the common-cause principle.

\section{ Basics of the branching space-time theory } 

Our formulation will use some of the very basic elements of the {\it 
Branching Space-time Theory } (BST). The 
aim of this theory was to solve the problem: How can we combine  
{\it 
relativity } and {\it indeterminism } in a rigorous theory? The 
underlying idea is that a true description of our world may require 
fusing Einstein spacetime with Prior/Thomason branching time. For 
further motivation and details of BST we refer to Belnap 1992. 

From BST we rely on the following postulates and definitions on the
primitives $\left\langle Our \: World, <\right\rangle $, where $Our\:
World$ 
is the totality of (possible) point events connected to us by a 
succession of causal paths, and $<$  
is an adaptation of the {\it causal ordering relation } among  
space-time 
points (the existence of a time-like or light-like path) to the domain  
of 
possible point events. We stress that no other primitives whatsoever 
figure in our discussion.

\paragraph{{\bf Postulate}} {\it ${\em Our\:  World}  $ is postulated  
to 
be a 
nonempty set; its members are called {\em point events}.  $<$  is 
postulated a dense partial order on $ Our\:  World $ such that every 
nonempty lower bounded chain has an infimum. 
A {\em history } is defined as a subset $h$  of $ Our\:  World $ 
maximal 
with respect to the property that if $h$  contains two 
point events, then it contains an upper bound for them. It is  
postulated 
that each nonempty upper bounded 
chain has a supremum in each history of which it is a subset. A {\em 
choice point } for two histories is defined 
as a point event maximal in their intersection. Given any nonempty 
lower bounded chain $E$  of point events 
in  $h_{1}-h_{2}$, it is postulated that there is at least one choice  
point 
$e_{1}$  for $h_{1}$  and $ h_{2}$ such that  $e_{1}<e_{2}$   for all 
$e_{2}\in E$ 
(this is the {\em ``prior choice principle''}). An {\em event } is  
defined 
as 
a nonempty set of point events.}

A good picture to have in mind is that each history is a space-time  
of 
General Relativity that is free from causal degeneracy (for example 
the Minkowski space-time). We do not use anything of the more 
detailed 
geometry of a space-time beyond its causal structure. So, if we  
wish, 
we 
can regard a history at a more abstract level as a causal space of  
point 
events satisfying the Kronheimer-Penrose (1967) axioms. 
The postulates above, especially the prior choice principle, constrain 
how these histories branch one from the other. 
The details are not, however, pertinent.
We add the following definitions for current purposes.
\begin{D}
\label{initial}
An {\em initial event }  $I$  is a nonempty upper bounded chain.  An 
{\em outcome event }  $ O $ is a nonempty lower bounded chain. A 
{\em stable event }  is both initial and outcome, and is contained in 
every history it overlaps.
\end{D}
Initial events represent situations that can have any one of a
variety of outcomes, which are represented by outcome events.   
Both
initial and outcome events have definite loci in $Our\: World$; but
the two sorts of events have different structures.

Initial events and outcome events, because of their respective  
structures, can fit
together into ``spreads.''  Roughly speaking, a spread -- which we  
are going
to define just below -- consists in a single
initial event, and a variety of individually possible but mutually
inconsistent outcomes of that initial. 
A spread with its outcomes is one way of representing an
``experiment'' and its possible outcomes. It also sometimes makes  
sense to think of a
spread as a choice-situation for an experimenter, with the outcomes
representing the available choices.
The meaning of a spread
is, however, more general; it can describe any kind of
indeterministic situation  without implying either
the presence or the absence of any human activity at the
``preparation'' of the initial event or ``observation'' of the
outcomes.
Thus the following formulation draws no distinction between
the ``a measurement process'' and any other process going on in Our
World:

\begin{D}
\label{spread}
A {\em spread } $\sigma $  is an ordered pair the first member of 
which is an initial event $ I $ and the 
second of which is a nonempty set  $\Omega $  of outcome events;  
we 
write  $\sigma =I\rightarrow \Omega $.  Each spread must satisfy 
the following three conditions:
\begin{itemize}
\begin{enumerate}
\item[i.] $ I $ causally precedes each  $O\in \Omega $.
\item[ii.]  Every history containing $ I $ overlaps some $O\in \Omega 
$.
\item[iii.] No history overlaps two distinct members of  $\Omega $.

\end{enumerate}
\end{itemize}
\end{D}

Thus given any historical course in which the initial event $ I $  
comes 
to a close, exactly one of the outcome 
events in  $\Omega $  commences. In all cases spreads
have two features: (i) they have a definite locus in Our World; and  
(ii) 
their internal structure has a definite causal structure, since their  
mutually
inconsistent ``outcomes'' are located after their initials in the causal  
ordering.

We next define an ``n-spread'' with its ``outcome vectors'' in order  
to
give us a convenient way of simultaneously considering
a number of experiments and their possible results.

\begin{D}
\label{nspread}
An {\em n-spread }  is a finite sequence of spreads: 
$$\Sigma =\left\{ 
\sigma _{i}=I_{i}\rightarrow \Omega _{i} \right\}_{ i=1,2,...,n }$$  
The collection of the initial events $\left\{ I_{1},I_{2}, ..., 
I_{n}\right\} $  is called the {\em set of initials } of $\Sigma $. An 
{\em outcome vector } ${\bf O } $ is 
a vector whose $i^{th}$   term is one of the outcomes of the $i^{th}$  
spread of  $\Sigma $,  ${\bf O}\in \Omega _{1}\times \Omega 
_{2}\times \cdots \times \Omega _{n}$.
\end{D}

The next definition uses the expressive power of BST to give an  
exact
account of what it means to say that a set of events is
``consistent.''
\begin{D}
\label{consistent}
A set of initial and outcome events is said to be {\em consistent }  if 
there is a history that 
contains each of its initial events and overlaps each of its outcome 
events. An outcome vector is said to be 
consistent if its terms form a consistent set of outcome events.
\end{D}
Observe that there is an important difference in how initial events
and outcome events enter into consistency-relations: In
the history in question, initial events
must finish, while outcome events must begin.  Given the difference  
in
their structures, that is to be expected.

We also distinguish several senses of ``consistency'' of an $ n
$-spread $\Sigma $.   
\begin{D}
\label{minconsistent}
$\Sigma $ is {\em minimally 
consistent } if its set of initials is consistent, and {\em maximally 
consistent } if all of its outcome vectors are consistent. And     
$\Sigma 
$ is {\em 1-consistent } if its set of initials is consistent with each 
single outcome of each of its 
spreads in the following sense: for each outcome event  $O\in  
\Omega 
_{i}$, where $1\leq i\leq n$, the set of events  $\left\{ I_{1},I_{2}, ..., 
I_{n}, O\right\} $
 is consistent.
\end{D}

Lastly, in contrast to the language available in most discussions of
Bell-like phenomena, the language of BST permits us to say with
absolute rigor what it is for a set of experimental situations to be 
spatially separated.

\begin{D}
\label{space-like}
$\Sigma $ is {\em space-like } iff it is minimally consistent, and no 
initial of one of its spreads 
causally precedes any outcome of a distinct spread.
\end{D}

The following sums up what will be used explicitly.

\newtheorem{F}{Fact}
\begin{F}
\label{Fact 1}
 Let $\sigma $  be a spread, let $\Sigma $  be an $n$-spread, and let  
$ 
E $ be an initial or outcome event of  $\Sigma $.
\begin{itemize}        
 \item $E$  is consistent with the initial of $\sigma $ if and only if $E$  
is consistent with some outcome of $\sigma $.
  \item $ E $ is consistent with the set of initials of an $ n$-spread if  
and only if $ E $ is consistent with some outcome
vector of $\Sigma $.
  \item If $ E $ is consistent with a set of initials and outcomes, then  
$ 
E $ is consistent with each element of the set.
  \item If $\Sigma $   is maximally consistent then it is 1-consistent. If 
it is 1-consistent then it is minimally consistent.
\end{itemize}
\end{F}

Using the definitions just given,
BST suggests the following partial
(and pre-probabilistic) analysis of the concept of the 
``common cause.'' The idea is that the 
question of a common cause does not arise unless you are given a 
spacelike $ n $-spread  $\Sigma $  that is 1-consistent, so that the  
joint 
coming  to a close of its initials is consistent with each outcome of  
each 
of its 
spreads. (One does not even look for a common cause if there is a
violation of 1-consistency; if, that is, an
otherwise
possible outcome of one the experiments is prevented merely by
initializing the other experiments.) But you are also given an  
outcome vector $ {\bf C } $ that is 
inconsistent -- in spite of the fact that each 
term of $ {\bf C } $ is consistent with the set of initials of  $\Sigma $. 
It is this combination that raises a question. And when 
the question does arise, one looks for a causal locus in the common 
past of the outcomes of spreads in  $\Sigma $. 
What one looks for can be understood in terms of Reichenbach's  
(1956) 
concept of ``screening off,'' here 
adapted to the language of consistency: The inconsistency of $ {\bf  
C } 
$ represents a ``correlation'' among the terms 
of the $ n $-spread, in spite of their space-like separation, and 
screening off makes the ``correlation'' disappear in the context of  
each 
outcome of the common cause $\sigma $. As a partial analysis, we 
collect the necessary 
conditions for a spread to be a common cause for an inconsistent 
outcome vector. If our analysis weren't 
restricted to inconsistency-type correlations, we would perhaps 
need to enrich or refine these conditions, but 
for present purposes the following definition works out:

\begin{D}
\label{common cause}
Let $\sigma $  be a spread,  $\Sigma $  a space-like $ n $-spread
that is 1-consistent, and $ 
{\bf C } $  an inconsistent outcome vector of  $\Sigma $. $\sigma $  is 
a common cause for $ {\bf C } $ in $\Sigma $   only if
\begin{itemize}
\item[CC1.] {\em (Causal priority)} The initial of $\sigma $  is 
causally prior to every outcome of every spread in $\Sigma $.
\item[CC2.] {\em (Consistency)} Each outcome of $\sigma $  is 
consistent with the set of initials of  $\Sigma $.
\item[CC3.] {\em (Screening off of inconsistency)} $\sigma $  
``screens 
off'' $ {\bf C } $. That is, each outcome $ O $ of $\sigma $  is 
inconsistent with 
some one term of $ {\bf C } $.

\end{itemize} \end{D}
The expression ``common cause for the
inconsistent outcome vector {\bf C}'' sounds a little awkward in
ordinary language; we use it as short for something like ``common
causal locus for {\bf C}'' or ``common-cause explanation for {\bf
C}''.  Whichever, it clearly and rigorously refers to an actual place
in $\it Our\: World$ such that what happens there ``explains away''
the 
space-like
correlation represented by {\bf C}.

\section{ BST formulation of the GHSZ-Mermin theorem}

We now turn to redescribing Mermin's version of the GHSZ theorem  
in 
the language of BST, dropping all 
language of particles, state, and systems. Nor do we even use much  
of 
the language of BST itself, confining ourselves to the notions 
introduced in the previous section. We shall, however, help  
ourselves 
to heuristic interpretations, with the caution that the hard  
information is 
contained in the explicit numbered stipulations below.

At each of the three ``stations'' $ 1,2,$  or $3$  there can be one of  
the 
two ``measurements'' of   $\sigma _{x}$   or $\sigma _{y}$   each of  
which can have one of two ``outcomes'' $-$  or $+$. We interpret  
this 
situation as follows. There are to begin 
with three widely separated ``pre-preparation'' initial events  $I_{1}, 
I_{2}$  and  $I_{3}$, with the interpretations e.g.
\begin{itemize}
\item[$I_{1}$:]Station $1$  has been pre-prepared (so to speak) so  
that 
it will be set to measure either on the $ x $ 
axis or on the $ y $ axis (but not both).
\end{itemize}
Next, the pre-preparation initial  $I_{1}$  (for example) has two 
possible outcomes  $x_{1}$ and $y_{1}$, with the interpretation e.g.
\begin{itemize}
\item[$x_{1}$:] Station $1$  is prepared to measure on the $x$   axis.

\end{itemize}
Now  $x_{1}$, for example, is not only an outcome event, but also  
an 
initial event, with its own two outcomes, $x^{-}_{1} $ 
 and  $x^{+}_{1} $, with the interpretation e.g.
\begin{itemize}
\item[$x^{-}_{1}$:] At station 1 the outcome was spin '$-$' on the $ x 
$ axis.
\end{itemize}
The picture to have is that the three stations are widely separated,  
but 
that at each station the events come 
in rapid succession. Keep in mind that our theoretical language  
implies 
little or nothing in the system/state vocabulary, much less in the 
language of classical physics, much less in the language of quantum 
mechanics. What happened happened; that's it. We record in the 
following ``stipulations'' the minimal story that quantum mechanics,  
no 
doubt, tells.

\newtheorem{ST}{Stipulation}

\begin{ST}
\label{1}
$\; $ 
\begin{itemize}
\item We use labels $\{1,2,3\}$  for the three ``stations,'' labels 
$\{x,y\}$  for the two ``measurement-types,'' and  labels $\{ -,+\}$   
for 
the two ``outcome-types.'' (The three quoted phrases are to be 
taken only as heuristic. Theoretically the labels we use are just  
labels; 
the suggested concepts do not figure as part of  theory.) We use $ i  
$ 
and $ j $ as ranging over $\{1,2,3\}$.
 
\item We shall consider three initial events $I_{i}$, six 
stable events $x_{i},y_{i}$, and twelve outcome 
events  $x^{-}_{i}, x^{+}_{i}, y^{-}_{i}, y^{+}_{i}$.
\item   These events fit together into spreads as follows. For $  
i=1,2,3$,
$\sigma _{i}, \sigma _{i}^{x}$, and $\sigma _{i}^{y}$ are spreads, 
where  
$\sigma _{i}=I_{i}\rightarrow \left\{ x_{i} y_{i}\right\} $  and 
$\sigma _{i}^{x}=x_{i}\rightarrow \left\{ x^{-}_{i} 
x^{+}_{i}\right\}$ and $\sigma _{i}^{y}=y_{i}\rightarrow 
\left\{ y^{-}_{i} y^{+}_{i}\right\}$. It follows from the 
stability of   $x_{i}$ and $y_{i}$ that for $ i=1,2,3$, $\sigma ^{\ast 
}_{i}=I_{i}\rightarrow \left\{ x^{-}_{i}\,  x^{+}_{i}\,  y^{-}_{i}\, 
y^{+}_{i} \right\}  $   is a spread.
\item   These spreads fit together into 3-spreads as follows: $\Sigma 
_{123}=\left\{ \sigma _{1}\sigma _{2}\sigma _{3}\right\} $, $\Sigma 
^{\ast }_{123}=\left\{ \sigma^{\ast } _{1}\sigma ^{\ast }_{2}\sigma 
^{\ast }_{3}\right\}$. We will also consider the following 3-spreads:
\begin{eqnarray}
\label{Sigmaspreads} 
\begin{array}{rcl} 
\Sigma _{xxx}& = & \left\{ \sigma ^{x} _{1}\sigma ^{x }_{2}\sigma 
^{x }_{3}\right\}\\
\Sigma  _{xxy}& =& \left\{ \sigma ^{x } _{1}\sigma ^{x }_{2}\sigma 
^{y }_{3}\right\}\\ 
\Sigma  _{xyx}& =& \left\{ \sigma ^{x} _{1}\sigma ^{y}_{2}\sigma 
^{x}_{3}\right\}\\ 
\Sigma  _{xyy}& =& \left\{ \sigma ^{x} _{1}\sigma ^{y}_{2}\sigma 
^{y}_{3}\right\}
\end{array} 
\end{eqnarray}
\end{itemize}
\end{ST}
	
That $\sigma _{i}$ and the $\sigma ^{x}_{i},\sigma ^{y}_{i} $ are
spreads already gives us the following spatio-temporal information:
for $ i=1,2,3$, we have $I_{i}< x_{i}, y_{i} < x^{-}_{i}, x^{+}_{i},
y^{-}_{i}, y^{+}_{i}$. The following stipulation is additional. 
\begin{ST}
\label{2}
The $3$-spread $\Sigma 
^{\ast }_{123}$   is space-like. This means that the stations are 
sufficiently far apart in a space-like sense, and the events at each 
station 
are sufficiently close in a time-like sense, so as to guarantee that no 
measurement outcome at one station has the initial of a distinct  
station 
in its causal past. A fortiori,  $\Sigma _{123}$  and all the $  
3$-spreads 
in (\ref {Sigmaspreads}) are spacelike.
\end{ST}

\begin{ST}
\label{3}
The following stipulations are based on quantum mechanics (see 
Section~\ref{sec-GHSZ}).
\begin{itemize}
\item An outcome vector of the 3-spread $\Sigma ^{\ast }_{123}$   
(and hence an outcome vector of a 3-spread in  
(\ref{Sigmaspreads})) is 
consistent if and only if (i) there is a {\em mixture } of $ x 
$ and $ y $ and an {\em even } number of minuses, or (ii) 
there is {\em no 
mixture } of x and y and an {\em odd } number of minuses. 
  \item Here are four examples, all used below. $\left( 
x^{+}_{1}x^{-}_{2} x^{+}_{3}\right) $, which is an outcome vector 
of  $\Sigma _{xxx}$, is consistent (no mixture of $ x $ and $ y $, odd 
$-$). Each of  $\left( x^{+}_{1}x^{-}_{2} y^{+}_{3}\right), \left( 
x^{+}_{1}y^{+}_{2} y^{-}_{3}\right)$  and  $\left( 
x^{+}_{1}y^{-}_{2} x^{+}_{3}\right)$  (an outcome vector of 
respectively $\Sigma _{xxy}, \Sigma _{xyy}$ and $\Sigma _{xyx}$), 
is inconsistent (mixed $ x $ and $ y $, odd number of  minuses).

  \item Furthermore, $\Sigma _{123}$ is maximally consistent: no
setting of measurement type at one station can be prohibited by any
selection of measurement types at the other stations. The same fact
can be restated by saying that all of the 3-spreads in question are
minimally consistent.  The 3-spreads in (\ref{Sigmaspreads}) are  
none
of them, however, maximally consistent. Instead these $3$-spreads
each has a property intermediate between minimal consistency and
maximal consistency: Each is 1-consistent, which is to say of each
that the joint realization of its measurement settings (initials)
permits the later occurrence of any single outcome of any of its
spreads (though of course not jointly!).

\end{itemize}
\end{ST}

An inconsistent outcome vector of a 1-consistent space-like  $ n 
$-spread is what sends us in search of a common cause. And maybe 
there is one. We do not prove that there isn't. Instead, we prove 
something weaker.
\begin{T}
\label{Th-NoCommon}
There is no single spread $\sigma $  that is a common cause for  
every 
inconsistent outcome vector of each of the four 3-spreads  $\Sigma 
_{xxx}, \Sigma _{xxy}, \Sigma _{xyy}$  and  $\Sigma _{xyx}$.
\end{T}
This theorem rules out, or seems to rule out, that the ``gedanken 
spin-conserving decay,'' if analyzed as an 
initial having a variety of possible outcomes, could serve as the 
single common-cause explanation of all the measurement  
correlations.
\paragraph{ Proof.} Suppose for {\em reductio } that $\sigma $  is a 
common cause of each inconsistent outcome vector of each of the 
3-spreads  $\Sigma _{xxx}, \Sigma _{xxy}, \Sigma _{xyy}$  and  
$\Sigma _{xyx}$. Therefore, even without a full analysis of the 
concept of a common cause, it is a part of our supposal that CC2  
and 
CC3 hold for  $\sigma $  and each of $\Sigma _{xxx}, \Sigma _{xxy}, 
\Sigma _{xyy}$  and  $\Sigma _{xyx}$.

We begin by choosing an outcome  $ O $ of $\sigma $. CC2 for 
$\Sigma _{xxx}$   implies that  $ O $ is consistent with the set of 
initials $\left\{ x_{1}x_{2}x_{3}\right\} $ of $\Sigma _{xxx}$. This 
implies by Fact \ref{Fact 1}  that  $ O $ is consistent with at least  
one 
consistent outcome vector of  $\Sigma _{xxx}$.
  At this point we ought to pick an {\em arbitrary } such consistent 
outcome vector, but instead we ask you to 
settle for a {\em persuasive example}: $\left( x^{+}_{1}x^{-}_{2} 
x^{+}_{3}\right) $, known to be consistent by Stipulation \ref{3}. So 
since  $ O $ is consistent with the outcome vector $\left( 
x^{+}_{1}x^{-}_{2} x^{+}_{3}\right) $, by Fact \ref{Fact 1}  
\begin{quote}
\begin{center} 
 $ O $ is consistent individually with each of $x^{+}_{1}, x^{-}_{2}$   
and $x^{+}_{3}$.
\end{center} 
\end{quote}
Now consider that $\left( x^{+}_{1} x^{-}_{2} y^{+}_{3}\right) $  
must by Stipulation \ref{3}  be inconsistent. So by CC3 for  $\Sigma 
_{xxy}$,  $ O $ must be inconsistent with one of those outcomes 
individually. But  $ O $ is consistent with each of the first two, so it 
must be that 
\begin{quote}
\begin{center} 
$ O $ is inconsistent with $y^{+}_{3}$.
\end{center} 
\end{quote}
CC2 for  $\Sigma _{xxy}$ implies that  $ O $ is consistent with the 
initials $\left\{ x_{1}x_{2}y_{3}\right\} $   of  $\Sigma _{xxy}$. By 
Fact \ref{Fact 1}  $ O $ is therefore consistent with the initial   
$y_{3}$  
of  $\sigma ^{y}_{3}$. The outcomes of  $\sigma ^{y}_{3}$   are 
$y^{+}_{3}$   and $y^{-}_{3}$. By Fact \ref{Fact 1}  $ O $ must be 
consistent with 
one of these two outcomes. Since  $ O $ is inconsistent with the 
former, it must be that
\begin{quote}
\begin{center} 
 $ O $ is consistent with $y^{-}_{3}$.
\end{center} 
\end{quote}
Next consider the outcome vector  $\left( 
x^{+}_{1}y^{+}_{2}y^{-}_{3}\right) $ of $\Sigma _{xyy}$. By 
Stipulation \ref{3}  it is inconsistent, and so by CC3 for $\Sigma 
_{xyy}$, together with previously established consistencies, it  
follows 
that
\begin{quote}
\begin{center} 
 $ O $ is inconsistent with  $y^{+}_{2}$.
\end{center} 
\end{quote}
Lastly, consider  $\left( x^{+}_{1}y^{-}_{2}x^{+}_{3}\right) $, 
which by Stipulation \ref{3}  is inconsistent. By CC3 for $\Sigma 
_{xyx}$, together with previously established consistencies, we have 
that
\begin{quote}
\begin{center} 
 $ O $ is inconsistent with  $y^{-}_{2}$.
\end{center} 
\end{quote}
But the inconsistency of  $ O $ with both  $y^{+}_{2}$  and 
$y^{-}_{2}$  implies, by Stipulation \ref{3}  and Fact \ref{Fact 1}, the 
inconsistency of  $ O $ with $y_{2}$. This contradicts the implication 
of CC2 for  $\Sigma _{xyx}$, by way of Fact \ref{Fact 1}, that  $ O $ 
is consistent with $y_{2}$, and concludes the {\em reductio }.

That of course was a proof by example. To generalize, recast the 
argument by replacing the definite labels with variables.

\hfill  $\Box $ 

\section{Limitations}

There seem to be three ways Our World might be:
\begin{itemize}

\item[Level I.] Deterministic, that is, there is only one history. 

\item[Level II.] Indeterministic, but without any ``strange"  
correlations 
between spatially separated happenings. In other words, each 
inconsistent outcome vector of each space-like n-spread has a 
common-cause explanation.

\item[Level III.] Indeterministic, and with ``strange" correlations 
between spatially separated happenings.

\end{itemize}

The strength of our result is that it establishes the relation between  
a
Bell-like phenomenon and a no-common-cause-like phenomenon  
with
absolute rigor, relying on the causal ordering in Our World as sole
primitive.  As we announced at the outset, however, there are
limitations.

\begin{itemize}
\item A frequent interpretation of the GHSZ story as well as the  
other 
Bell-like theorems is that they
show that certain phenomena predicted by quantum mechanics (but
describable pre-theoretically, without quantum mechanics) are
{\it incompatible with determinism}. One of us has elsewhere 
questioned the legitimacy of this
interpretation as arising from insufficient care in applying the
usual formalism (Szab\'o 1995a,b). Certainly our proof cannot bear
such an interpretation, because it begins by {\em stipulating} the
existence of certain spreads, which are explicitly indeterministic
phenomena.  We may in fact be permitted to doubt that other proofs
fare better in this respect.  Although these proofs are not
sufficiently rigorous to be sure, they seem to share with our proof
the {\em hypothesis} that indeterministic phenomena occur.  

\item Our proof of Theorem \ref{Th-NoCommon} makes no use of  
the
causal-priority condition CC1 of Definition \ref{common cause}.  
More
work is needed in understanding the interplay of the idea of causal
priority with other concepts.  We need especially to advance our
understanding of the proposition that the values of certain
measurements cannot be arranged in advance.

\item Bell-like theorems are often interpreted as showing that
certain phenomena predicted by quantum mechanics are inconsistent
with the principle of the common cause; or, in other words, that
these phenomena involve space-like correlations without a common
cause. We are in this paper explicit that Theorem
\ref{Th-NoCommon} does not say, simply, that there are correlations
without a common cause.  Instead, it says what it can prove: There
are certain sets of correlations such that one cannot find a single
common-cause locus for all of them.  In other words, although
Theorem \ref{Th-NoCommon} does not strictly imply that Our World
belongs to Level III, it moves us sharply in that direction.
So even though our result succeeds
in ruling out that a single ``gedanken decay'' can account for all
the correlations, its limited nature suggests the interest in
finding a better theorem.

\end{itemize}

\section*{References}
\begin{list}%
{ }{\setlength{\itemindent}{-15pt}
\setlength{\leftmargin}{15pt}}

\item Belnap, N. (1992): ''Branching space-time,'' {\em Synthese}, 92, 
385-434.
\item Bohm, D. and Hiley, B. J. (1993): {\em The Undivided 
Universe}, Routledge, London and NY.
\item Brown, H. R. and Svetlichny, G. (1990): {\em Found. Phys.} {\bf 
20}, 1379-1387.
\item Greenberger, D. M., Horn, M. A., Shimony, A. and Zeilinger, A. 
(1990): ``Bell's theorem without inequalities.'', {\em Am. J. Phys.} {\bf 
58}, pp. 1131-1143.
\item Kronheimer, E. H. and Penrose, R. (1967): ``On the structure of 
causal spaces,'' {\em Proceedings of the Cambridge Philosophical 
Society}, {\bf 63}, 418. 
\item Mermin, N. D. (1990a): ``What's wrong with these elements of 
reality?'', {\em Physics Today}, June.
\item Mermin, N. D. (1990b): ``Simple Unified Form for the Major 
No-Hidden-Variables Theorem", {\em Phys. Rev. Lett.}, {\bf 65}, 
3373-3376.   
\item Reichenbach, H. (1956):{\em Direction of Time}, University of 
California Press, Berkeley.
\item Szab\'o, L. E. (1995a): ``Is quantum mechanics compatible with  
a deterministic universe? Two interpretations of quantum  
probabilities" {\it Foundations of Physics Letters}, {\bf 8}, 421-440.
 
\item Szab\'o, L. E. (1995b): ``Quantum mechanics in an entirely 
deterministic universe" {\it Int. J. Theor. Phys.}, {\bf 34}, 1751-1766. 

\end{list}

\end{document}